\documentclass[aps,prb, reprint, superscriptaddress,preprintnumbers,amssymb]{revtex4-2}
\usepackage{graphicx}
\usepackage{dcolumn}
\usepackage{bm}

\begin{document}

\preprint{ver3.1}

\title{Apparent nonreciprocal transport in FeSe bulk crystals}



\author{Taichi Terashima}
\email{TERASHIMA.Taichi@nims.go.jp}
\affiliation{Research Center for Materials Nanoarchitectonics (MANA), National Institute for Materials Science, Tsukuba 305-0003, Japan}
\author{Shinya Uji}
\affiliation{Research Center for Materials Nanoarchitectonics (MANA), National Institute for Materials Science, Tsukuba 305-0003, Japan}
\author{Yuji Matsuda}
\affiliation{Department of Physics, Kyoto University, Kyoto 606-8502, Japan}
\author{Takasada Shibauchi}
\affiliation{Department of Advanced Materials Science, University of Tokyo, Kashiwa, Chiba 277-8561, Japan}
\author{Shigeru Kasahara}
\email{kasa@okayama-u.ac.jp}
\affiliation{Research Institute for Interdisciplinary Science, Okayama University, Okayama 700-8530, Japan}


\date{\today}

\begin{abstract}
We performed low-frequency ac first- and second-harmonic resistance measurements and dc $I-V$ measurements on bulk FeSe crystals in a temperature range between 1.8 and 150 K and in magnetic field up to 14 T.
We observed considerable second-harmonic resistance, indicative of nonreciprocal charge transport, in some samples.
By examining correlation between contact resistances and second-harmonic signals, we concluded that the second-harmonic resistance was not due to the genuine nonreciprocal transport effect but was caused by joule heating at a current contact through the thermoelectric effect.
Our conclusion is consistent with a recent preprint (Nagata \textit{et al.}, arXiv:2409.01715), in which the authors reported a zero-field superconducting diode effect in devices fabricated with FeSe flakes and attributed it to the thermoelectric effect.
\end{abstract}


\maketitle



%

\section{Introduction}
Nonreciprocal charge transport in quantum materials lacking space-inversion symmetry attracts considerable attention in recent years, not only because of fundamental interest but also because of potential technological importance \cite{Ideue21ARCMP}.
The nonreciprocal charge transport here refers to a state where the electrical resistance differs between a right-going ($+I$) and a left-going current ($-I$).
It usually requires that time-reversal symmetry is also broken.
This is because, as long as the time-reversal symmetry is present, Kramers degeneracy enforces the relation $\epsilon_{\sigma}(k) = \epsilon_{-\sigma}(-k)$ on the electronic band energy, which is usually considered to preclude the nonreciprocal charge transport.
Some recent theories however claim that the nonreciprocal transport can occur without time-reversal symmetry breaking by considering skew scattering \cite{Isobe20SciAdv} or electron correlation \cite{Morimoto18SciRep}, for example.

The most impressive manifestation of the nonreciprocal charge transport may be a superconducting diode where the resistance is zero for one current direction while finite for the other.
Although the initial realization of the superconducting diode with an artificial superlattice [Nb/V/Ta]$_n$ required the application of a magnetic field to break time-reversal symmetry \cite{Ando20Nature}, there are recent reports that superconducting diodes may be achieved at zero magnetic field without breaking time-reversal symmetry \cite{Liu24SciAdv, Nagata24condmat}.
Especially intriguing is the observation of the superconducting diode effect at zero field in devices fabricated with an FeSe flake \cite{Nagata24condmat}.
FeSe is a unique parent compound of iron-based superconductors \cite{Hsu08PNAS}:
It exhibits a tetragonal-to-orthorhombic structural phase transition at $T_s \sim 90$ K but no antiferromagnetic order.
It becomes superconducting below about $T_c \sim 9$ K.
The space group of the low-temperature orthorhombic lattice is $Cmme$ preserving inversion symmetry \cite{Margadonna08ChemComm, Rossler22PRB}.
Hence the superconducting diode effect was observed despite that both space-inversion and time-reversal symmetries were retained.
The authors argued that the observed effect was actually due to the large thermoelectric effect of FeSe rather than the intrinsic nonreciprocal transport effect \cite{Nagata24condmat, Kasahara16NatCommun}.
Namely, because the shapes of the FeSe flakes were asymmetric, roughly triangular, the two current contacts attached to them differed in size and hence the joule heating was larger at the narrower contact, resulting in the temperature gradient.
This caused additional unidirectional current flow via the large thermoelectric effect and broke superconductivity for one current direction.
In this work, we show that similar apparent nonreciprocal charge transport is observed even in bulk FeSe crystals.

\begin{figure}
\includegraphics[width=8.6cm]{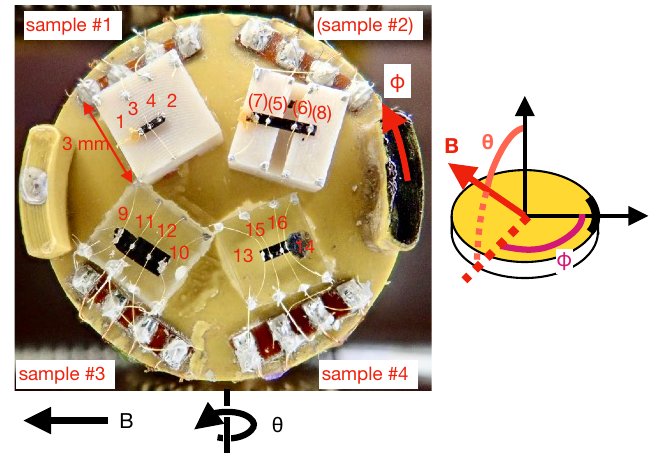}
\caption{\label{Sample}FeSe samples mounted on a rotation platform at $\phi$ = 0 and $\theta$ = $-90^{\circ}$.
Contact numbers are indicated.
The polar $\theta$ and azimuthal $\phi$ angles of the applied magnetic field were defined with respect to the platform (right panel).
$\theta = 0$ corresponds to $B \parallel c$.
Data from sample \#2 were not used because the electrical contacts were unstable.
The dimensions of the samples are 1.0 $\times$ 0.26 $\times$ 0.12 mm$^3$, 2.2 $\times$ 0.78 $\times$ 0.02 mm$^3$, and 1.0 $\times$ 0.40 $\times$ 0.02 mm$^3$ for \#1, 3, and 4, respectively.}
\end{figure}

\begin{table}
\caption{\label{Tab} Contact resistance.  The accuracy is estimated to be about 0.1 $\Omega$.
}
\begin{ruledtabular}
\begin{tabular}{lllllllll}
sample &  \multicolumn{8}{l}{(contact number) contact resistance in $\Omega$} \\
\hline
\#1 & (1) & 0.7 & (2) & 4.7 & (3) & 0.7 & (4) & 0.4\\
\#3 & (9) & 0.3 & (10) & 0.7 & (11) & 0.4 & (12) & 0.6\\
\#4 & (13) & 0.6 & (14) & 3.0 & (15) & 0.5 & (16) & 0.8\\
\end{tabular}
\end{ruledtabular}
\end{table}

\section{Experiments}
High-quality single crystals of FeSe were grown by a chemical vapor transport method \cite{Bohmer13PRB}.
Figure 1 shows the experimental setup.
We chose approximately rectangular-shaped single crystals, avoiding multiple fused crystals.
Their length directions were along the tetragonal [100] axis, and their dimensions are given in the caption.
Four 10-$\mu$m gold wires were spot-welded to each sample and the contacts were reinforced by conducting silver paint.
The contacts were numbered as shown in the figure.
Although four samples were prepared, we do not mention data from sample \#2 because the electrical contacts were unstable.
The residual resistivity ratio RRR at $T$ = 11 K and the midpoint $T_c$ are (RRR, $T_c$) = (20, 8.5 K), (26, 9.1 K), and (29, 9.1 K) for sample \#1, 3, and 4, respectively.
The contact resistances (except those for sample \#2) were estimated at room temperature from two-wire measurements through different contacts and wire resistances estimated from their lengths and resistivities with an accuracy of about 0.1 $\Omega$ (Table I).
Each sample except sample \#3 was fixed on a polyether ether ketone (PEEK) or quartz substrate at one end with epoxy or GE (General Electric) varnish to avoid vibration due to Lorentz force acting on ac current.
Sample \#3 was fixed on an FRP (fiber reinforced plastic) substrate with vacuum grease.
The substrate material and glue were different between the samples, but this was not intentional, simply because some of the samples had previously been used for different purposes and were reused in this experiment. 
The samples were mounted on a two-axis rotation platform and loaded into a He-4 variable temperature insert equipped with a 17-T superconducting magnet.
The polar $\theta$ and azimuthal $\phi$ angles of the applied magnetic field were defined with respect to the platform as shown in the figure.
We performed low-frequency ($f$ = 7--19 Hz) ac lock-in measurements of the resistance at the fundamental and second-harmonic frequencies and dc $I-V$ measurements.

\section{Nonreciprocal transport and nonlinear $I-V$ characteristics}
Before presenting experimental data, we briefly review the nonreciprocal transport and nonlinear $I-V$ characteristics.

Rikken \textit{et al.} provided a basic framework for the nonreciprocal transport under broken time-reversal symmetry on the basis of symmetry arguments \cite{Rikken01PRL,Rikken05PRL}.
They showed that the nonreciprocal transport appears in chiral and polar structures, and the resistance depends linearly on both the electrical current and the magnetic field as follows:
\begin{equation}
R = R_o \left(1 + \beta B^2 + \gamma \bm I \cdot  \bm B \right)
\end{equation}
for chiral structures and
\begin{equation}
R = R_o \left[1 + \beta B^2 + \gamma \bm I \cdot \left(\bm P \times \bm B \right) \right],
\end{equation}
for polar structures.
Here, $R_o$ is the resistance without magnetic field, $\beta B^2$ the normal quadratic magnetoresistance, $\gamma$ the magnitude of the nonreciprocal effect, and $\bm P$ the axis of the polar structure. 
Because $\gamma$ depends on the sample size, a normalized parameter $\gamma^{\prime} = A\gamma$, where $A$ is the cross-sectional area of a sample, sometimes used for comparison between different materials \cite{Ideue21ARCMP}.

The nonreciprocal transport under time-reversal symmetry is more elusive.
Theoretical studies suggest that it requires electron correlation or skew scattering \cite{Morimoto18SciRep, Isobe20SciAdv}.

We now derive the expressions that will be used in analyzing experimental data.
We assume the following $I-V$ relation:
\begin{equation}
V = R_1 I + R_2 I^2 + R_3 I^3,
\end{equation}
where $R_i$ is a function of temperature $T$ and magnetic field $B$.

For dc $I-V$ measurements, we perform measurements at a positive ($+B$) and a negative ($-B$) field and calculate the symmetric (wrt $B$) $R^s$ and the antisymmetric component $R^a$ of the resistance:
\begin{eqnarray}
R^s(B) &=& \frac{1}{2}\frac{V(B)+V(-B)}{I} \nonumber \\
&=& R_1^s(B)+R_2^s(B)I+R_3^s(B)I^2, \mathrm{and} \\
R^a(B) &=& \frac{1}{2}\frac{V(B)-V(-B)}{I} \nonumber \\
&=& R_1^a(B)+R_2^a(B)I,
\end{eqnarray}
where we have neglected the antisymmetric part $R_3^a$.

For ac lock-in measurements, we apply the current $I = I_o \sin \omega t$ where $\omega = 2\pi f$ and obtain the following relations:
\begin{eqnarray}
V_1 &=& R_1 I = R_1 I_o \sin \omega t, \textrm{and}\\
V_2 &=& R_2 I^2 = R_2 I_o^2 \sin^2 \omega t \nonumber \\
&=& \frac{1}{2} R_2 I_o^2 \left[1-\sin\left(2\omega t + \frac{\pi}{2}\right)\right].
\end{eqnarray}
Because lock-in amplifiers output rms values, their outputs are expressed as follows:
\begin{eqnarray}
V_1^{Lx} &=& \frac{1}{\sqrt{2}}R_1I_o = R_1 I_{rms}, \mathrm{and} \\
V_2^{Ly} &=& -\frac{1}{2\sqrt{2}}R_2 I_o^2 =  -\frac{1}{\sqrt{2}}R_2 I_{rms}^2,
\end{eqnarray}
where $V_i^{Lx(y)}$ is the in-phase (quadrature) component of the $i$-th harmonic lock-in voltage.
The (anti)symmetric part $R_i^{s(a)}$ ($i=1,2$) is obtained by performing the measurements at $+B$ and $-B$ and (anti)symmetrizing $R_i$.

$R_1^s$ is the usual longitudinal resistance.
$R_2^s$ and $R_2^a$ can be due to genuine nonreciprocal transport effects under time-reversal symmetry and under broken time-reversal symmetry, respectively.
However, in addition to those intrinsic effects, we have to consider extrinsic origins (or experimental artifacts) as well.
$R_1^a$ can arise from the contamination of the Hall voltage.
$R_2^s$ and $R_2^a$ can arise from the Seebeck and Nernst effect, respectively:
a temperature gradient due to joule heating at contacts, which is proportional to $I^2$, gives rise to a voltage proportional to $I^2$ due to the Seebeck and Nernst effects.
Notice that the Nernst effect is antisymmetric with respect to $B$.
$R_3^s$ can arise from a resistance change $\Delta R$ due to sample heating:
a temperature change $\Delta T \propto I^2$ due to sample heating gives rise to $\Delta R = (\mathrm{d}R/\mathrm{d}T) \Delta T$ and hence a voltage proportional to $I^3$. 

\section{Results}
\subsection{sample \#4}

\begin{figure}
\includegraphics[width=8.6cm]{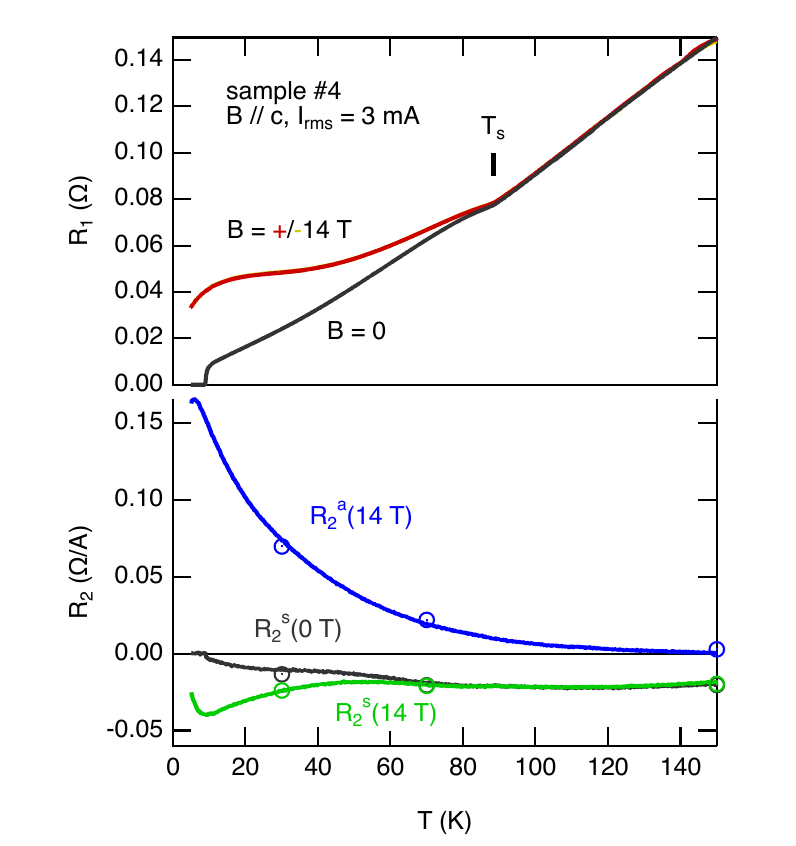}
\caption{\label{Sample4}Ac resistance versus temperature measurements on sample \#4 at $B$ = 0 and $\pm$14 T along the $c$ axis.
The contacts 13 and 14 were used as current contacts (see Fig. 1, contact configuration N).
For the second-harmonic resistance, the symmetric $R_2^s$ and the antisymmetric part $R_2^a$ at $B$ = 14 T and $R_2$ = $R_2^s$ at $B$ = 0 are shown.
The circles show values deduced from dc $I-V$ measurements (ex. Fig. 5) for comparison (after corrected for the field-angle difference, if necessary).
}
\end{figure}

Figure 2 shows results of ac resistance versus temperature measurements on sample \#4 at $B$ = 0 and $\pm$14 T applied along the $c$ axis ($\theta$ = 0).
The contacts 13 and 14 were used as current contacts (see Fig. 1, hereafter referred to as contact configuration N, N indicating `normal'), and the current flows in the $ab$ plane.
The frequency and magnitude of the current were $f$ = 18.8 Hz, and $I_{rms}$ = 3 mA.
Note that the contact resistance of 14 is relatively large, being 3.0 $\Omega$ (Table I).
For the second-harmonic resistance $R_2$, the symmetric part $R_2^s$( = $R_2$) at $B$ = 0 as well as the symmetric $R_2^s$ and the antisymmetric part $R_2^a$ at $B$ = 14 T are shown.
The symmetric part $R_2^s$ at $B$ = 0 is already finite at 150 K, shows a relatively weak temperature dependence, and its magnitude starts decreasing near 80 K.
The symmetric part $R_2^s$ at $B$ = 14 T is indistinguishable from that at $B$ = 0 above $T_s$, deviates downward below $T_s$, and starts approaching zero below $\sim$9 K.
The two curves exhibit a very faint kink at $T_s$.
The antisymmetric component $R_2^a$ at $B$ = 14 T appears from slightly above $T_s$ and increases with decreasing temperature down to $\sim$6 K, then starts decreasing.
No clear anomaly was observed in $R_2^a$ at $T_s$.
The parameter $\gamma^{\prime}$ is 2.6$\times 10^{-9}$ m$^2$A$^{-1}$T$^{-1}$ at 6 K (we re-defined $\gamma$ as $\gamma = R_2^a(B)/[R_1^s(B) B]$).
Although it is smaller than $\gamma^{\prime} \sim$10$^{-7}$ m$^2$A$^{-1}$T$^{-1}$ in WTe$_2$ and ZrTe$_5$ \cite{Yokouchi23PRL, Wang22PRL}, it is considerably larger than $\sim$10$^{-12}$ m$^2$A$^{-1}$T$^{-1}$ in BiTeBr \cite{Ideue17NatPhys}.

\begin{figure}
\includegraphics[width=8.6cm]{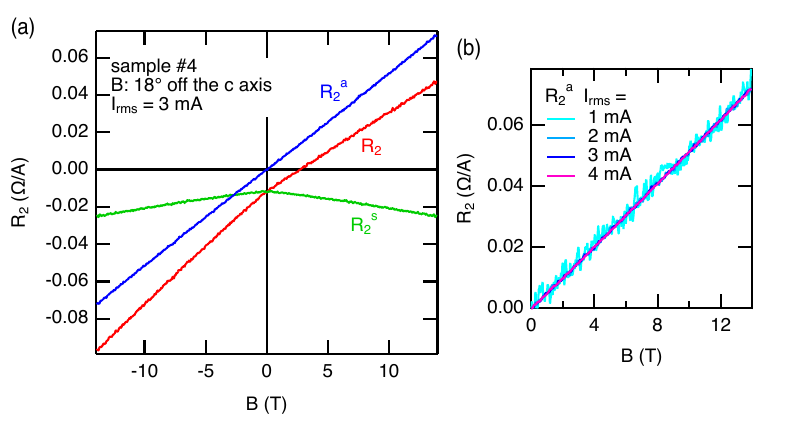}
\caption{\label{Sample}Second-harmonic resistance versus magnetic field at $T$ = 30 K for sample \#4 with contact configuration N.
The magnetic field was applied at an angle $\theta$ of 18$^{\circ}$ from the $c$ axis ($\phi$ = 90$^{\circ}$).
(a) The raw data $R_2$ and its decomposition into the symmetric $R_2^s$ and the antisymmetric part $R_2^a$ are shown.
(b) The antisymmetric part $R_2^a$ for four different current values.
}
\end{figure}

Figure 3(a) shows the magnetic field dependence of the second-harmonic resistance $R_2$ and its decomposition into the symmetric $R_2^s$ and the antisymmetric part $R_2^a$.
The magnetic field was applied at an angle $\theta$ of 18$^{\circ}$ from the $c$ axis ($\phi$ = 90$^{\circ}$) for a technical reason.
The antisymmetric part is linear in $B$.
Figure 3(b) shows the antisymmetric part $R_2^a$ measured with different current strengths.
All the curves coincide within error, indicating that the resistance linearly depends on the current.
Thus the bilinear nature (wrt $B$ and $I$) of the nonreciprocal transport is confirmed.

\begin{figure}
\includegraphics[width=5.5cm]{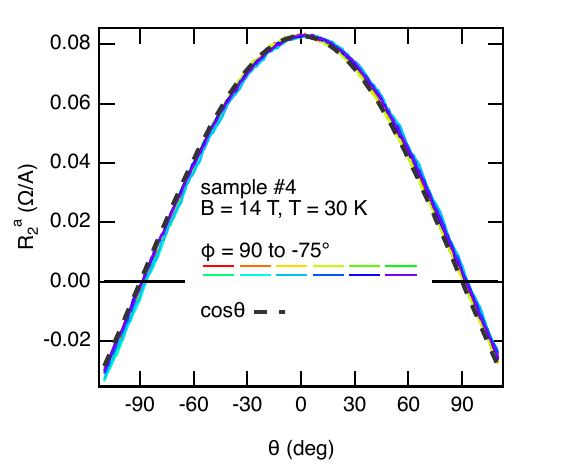}
\caption{\label{Sample}Dependence of $R_2^a$ in sample \#4 with contact configuration N at $B$ = 14 T and $T$ = 30 K on the polar angle $\theta$ of the applied field measured at different $\phi$'s.
The azimuthal angle $\phi$ was varied from 90 to -75$^{\circ}$ in steps of 15$^{\circ}$. 
The thick broken curve shows a $\cos \theta$ dependence.
}
\end{figure}

Figure 4 shows the field-angle $\theta$ dependence of the antisymmetric part $R_2^a$ at $B$ = 14 T and $T$ = 30 K.
The curves for different $\phi$'s coincide and follow the $\cos \theta$ dependence indicated by the thick broken curve within error.
This is the behavior expected from Eq. (2) when the vector $\bm P$ is in the $ab$ plane.

\begin{figure}
\includegraphics[width=8.6cm]{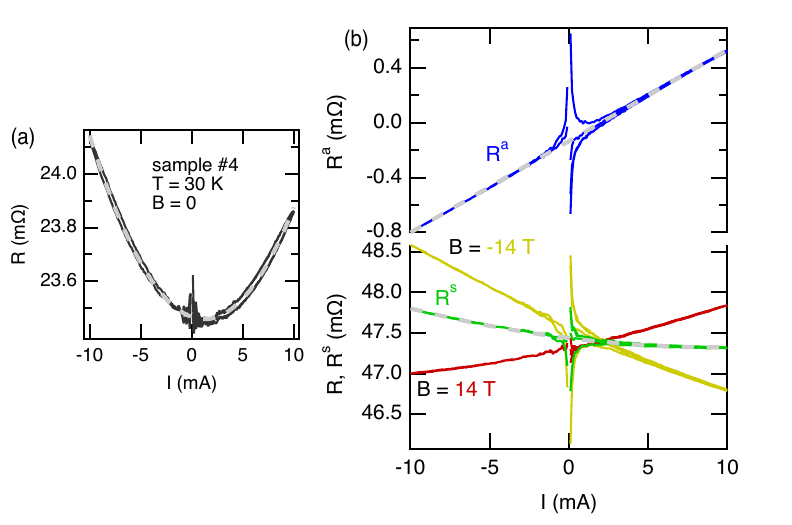}
\caption{\label{Sample}Dc $I-V$ measurements on sample \#4 with contact configuration N at $T$ = 30 K and (a) at $B$ = 0 and (b) $\pm$14 T plotted in the form of $R$ versus $I$.
The magnetic field direction is the same as that in Fig. 3: $(\theta, \phi) = (18^{\circ}, 90^{\circ})$.
In (b), the symmetric (wrt $B$) $R^s$ and the antisymmetric component $R^a$ are also shown.
Second-order polynomial fits to $R(I)$ at $B$ = 0 (a) and $R^s(I)$ at $B$ = 14 T (b) and a linear fit to $R^a(I)$ at $B$ = 14 T (b) are shown by grey broken lines.
At low current values (i.e., $|I| < \sim2$ mA), measured voltages were too small to accurately determine $R$, which caused apparently diverging behavior of $R$ as $I \rightarrow 0$.
}
\end{figure}

We now turn to dc $I-V$ measurements.
For those measurements, the current was swept as follows: 0 $\to$ 10~mA $\to$ 0 $\to$ $-$10~mA $\to$ 0 $\to$ 10~mA $\to$ 0.
Figures 5(a) and (b) show results of dc $I-V$ measurements on sample \#4 with contact configuration N at $T$ = 30 K for $B$ = 0 and $\pm$14 T, respectively.
The magnetic field direction is the same as that in Fig. 3: i.e., $(\theta, \phi) = (18^{\circ}, 90^{\circ})$.
The measured $R(I)$ (= $R^s$) curve at $B$ = 0 is asymmetric with respect to $I$ = 0, indicating a finite linear-in-$I$ term [Fig. 5(a)].
A fit to Eq. (4) (broken line) gives a linear coefficient $R_2^s$ = -0.013 $\Omega$/A.
It is in good agreement with ac measurements [see circles in Fig. 2 and $R_2^s$ at $B$ = 0 in Fig. 3(a)].
The existence of the quadratic term suggests joule heating of the sample.
Using the fit coefficient $R_3^s$ = 5.4 $\Omega$/A$^2$ and d$R$/d$T$  = 0.81 m$\Omega$/K from Fig. 2, we can estimate the increase in the \textit{average} sample temperature to be 0.06 K at $I$ = 3 mA.
Similarly, we fit Eq. (4) to the symmetric component $R^s$ at $B$ = 14 T [Fig. 5(b)] and find that $R_2^s$ = -0.024 $\Omega$/A and that $R_3^s$ = 1.3 $\Omega$/A$^2$.
The former is again in good agreement with ac measurements [Fig. 2 and Fig. 3(a)], and the latter gives an estimate of the average sample temperature increase of 0.08 K at $I$ = 3 mA (d$R$/d$T$  = 0.15 m$\Omega$/K at $B$ = 14 T from Fig. 2).
The antisymmetric component $R^a$ at $B$ = 14 T is nicely linear in $I$ (except at very low current values where voltage is too small to accurately determine $R$), suggesting the existence of the nonreciprocal transport under broken time-reversal symmetry [Fig. 5(b)].
By fitting Eq. (5) to the $R^a$ curve, we obtain $R_2^a$ = 0.066 $\Omega$/A, which is in good agreement with ac measurements [Fig. 2 and Fig. 3(a)].
Notice that the fit line intersects the $I$ = 0 axis at a finite value, i.e., $R_1^a \neq 0$.
This indicates that the measured voltage was contaminated by the transverse voltage due to the Hall effect, and hence that the transverse voltage due to the Nernst effect could also be detected with this contact arrangement.

\begin{figure}
\includegraphics[width=5.5cm]{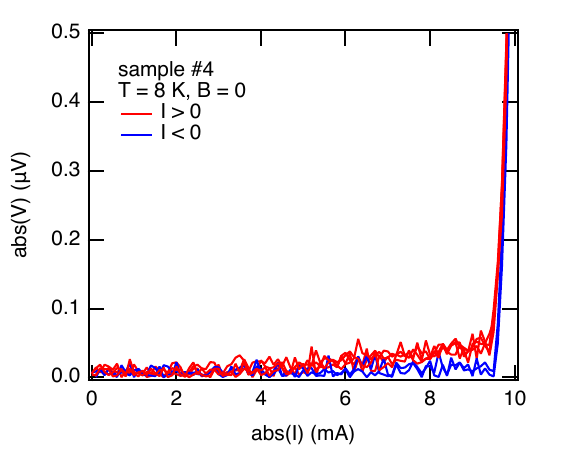}
\caption{\label{Sample}Absolute voltage versus absolute current in sample \#4 measured at $T$ = 8 K without magnetic field.
}
\end{figure}

Finally, we show that a faint field-free superconducting diode effect was observed in bulk FeSe.
Figure 6 plots the absolute voltage versus absolute current measured at $T$ = 8 K without a magnetic field.
For a current range roughly from 7 to 9 mA, the voltage is almost zero for negative current, while it is finite for positive current.
The observed effect is much weaker than reported in \cite{Nagata24condmat}.
The gradual rise in the voltage for positive current before the sharp increase above $I$ = 9.5 mA suggests the existence of areas where $T_c$ and the critical current density $J_c$ are smaller than in other areas.
Probably, in those areas, the thermally-induced current can become a significant fraction of the critical current, as in the tiny FeSe flakes used in \cite{Nagata24condmat}, giving rise to the superconducting diode effect.

\begin{figure}
\includegraphics[width=5.5cm]{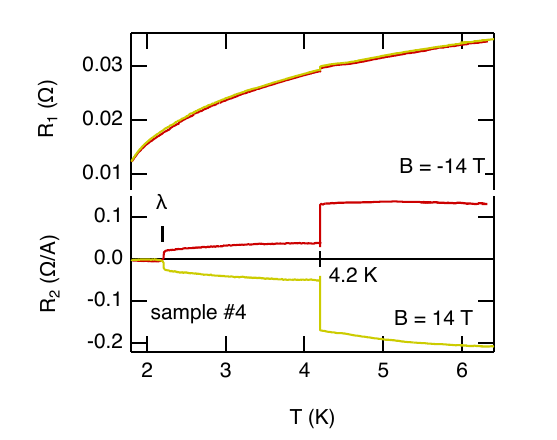}
\caption{\label{Sample}First- and second-harmonic resistance ($R_1$ and $R_2$) measured on sample \#4 with contact configuration N at $B$ = $\pm$14 T as a function of temperature.
}
\end{figure}

The data presented above strongly suggested the existence of the nonreciprocal transport coefficient $R_2^s$ under time-reversal symmetry as well as $R_2^a$ under broken time-reversal symmetry that is compatible with Eq. (2).
However, we were forced to reconsider the origin of the nonzero $R_2^s$ and $R_2^a$ by the following observation:
Figure 7 shows the first- and second-harmonic resistance measured on sample \#4 with contact configuration N at $B$ = $\pm$14 T as a function of temperature.
Initially, the sample was immersed in superfluid $^4$He at $T$ = 1.8 K, when the first-harmonic resistance $R_1$ was finite, but the second-harmonic one $R_2$ was almost zero.
As the temperature was slowly raised, the second-harmonic resistance $R_2$ suddenly jumped at $T$ = 2.2 K, i.e., the $\lambda$ point, and became clearly finite. 
With further increasing temperature, it again jumped and became much larger at $T$ = 4.2 K as the sample became no longer surrounded by liquid helium.
The first-harmonic resistance $R_1$ also showed a slight kink at $T$ = 4.2 K.
These observations strongly suggested that joule heating at a current contact (or maybe both current contacts) and resulting temperature gradient in the sample caused the nonzero $R_2^s$ and $R_2^a$ through the large thermoelectric effect \cite{Kasahara16NatCommun}.
Heat exchange between the sample and the environment and hence the temperature gradient inside the sample depend on whether the sample is surrounded by liquid helium and also whether the liquid is superfluid or not.
Superfluid helium-4 is expected to provide the best heat exchange, and hence the least temperature gradient is expected, which is consistent with the vanishingly small $R_2$ below the $\lambda$ point (Fig. 7).
In the following, we substantiate this conjecture.

\begin{figure}
\includegraphics[width=5.5cm]{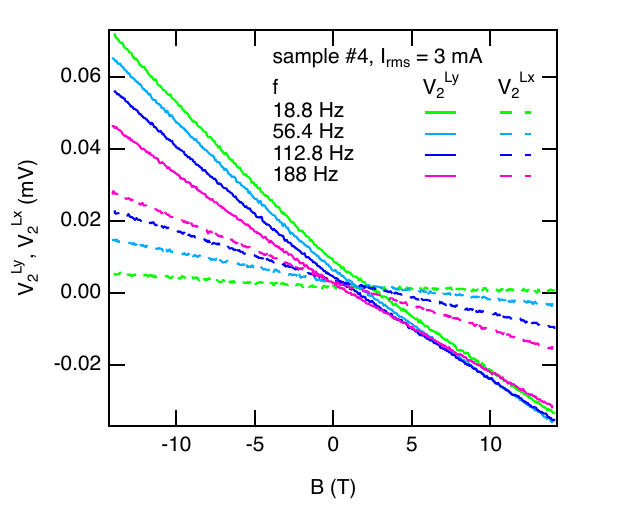}
\caption{\label{Sample}Quadrature and in-phase second-harmonic voltage ($V_2^{Ly}$ and $V_2^{Lx}$) measured on sample \#4 with contact configuration N at $T$ = 30 K for different current frequencies.
}
\end{figure}

Figure 8 shows the second-harmonic voltage as a function of field measured on sample \#4 at $T$ = 30 K with $I_{rms}$ = 3 mA.
Four different frequencies were used, and both the quadrature $V_2^{Ly}$ and the in-phase component $V_2^{Lx}$ are shown.
Note that the former $V_2^{Ly}$ corresponds to $R_2$.
As the frequency increases, the magnitude of $V_2^{Ly}$ decreases, and $V_2^{Lx}$ develops, indicating increasing phase delay.
This is expected behavior when the second-harmonic voltage is due to the thermoelectric effect:
When an ac current $i \sin \omega t$ is applied, the power dissipation at a current contact with a resistance $r$ is $(1/2)i^2r[1-\sin(2\omega t + \pi/2)]$.
Accordingly, the temperature difference between the contact and the environment oscillates as $\sin(2\omega t + \pi/2-\varphi)$, in addition to a constant increase \cite{Dubson89PRB, Lu01RSI}.
The phase delay $\varphi$ increases with the frequency, resulting in the increase in the phase-shifted component $V_2^{Lx}$.

\begin{figure}
\includegraphics[width=8.6cm]{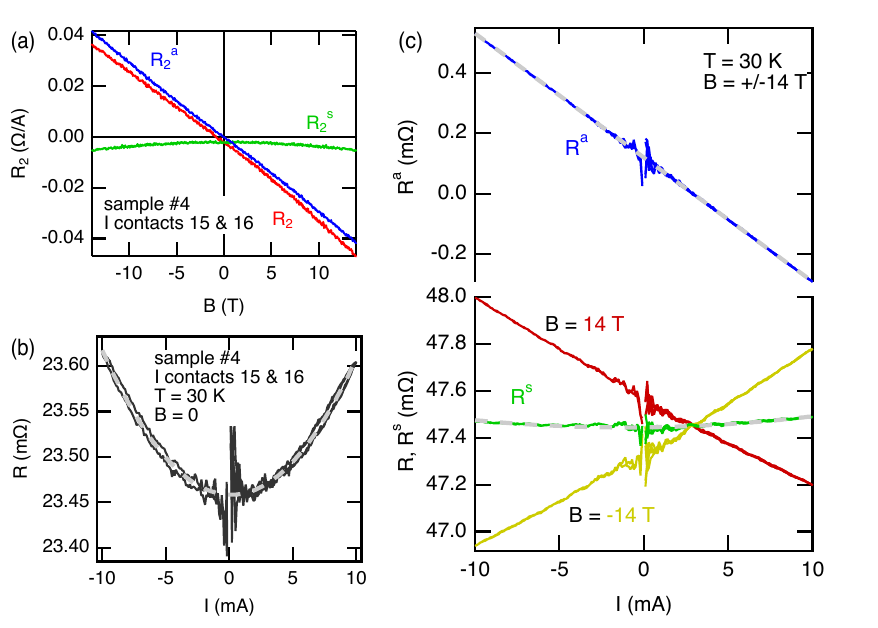}
\caption{\label{Sample}Sample \#4 with contact configuration R: contacts 15 and 16 were used as current contacts.  $T$ = 30 K.  (a) Second-harmonic resistance $R_2$ as a function of magnetic field.  The decomposition into $R_2^s$ and $R_2^a$ is shown.  (b) Resistance versus current at $B$ = 0.  (c)  Resistance versus current at $B$ = $\pm$14 T.  The symmetric (wrt $B$) $R^s$ and the antisymmetric part $R^a$ are also shown.  The grey broken lines show fitting results.
}
\end{figure}

All the data presented in Figs. 2--8 were obtained by using contacts 13 and 14 as current contacts, and 15 and 16 as voltage ones (namely, contact configuration N).
We also measured the resistance using contacts 15 and 16 as current contacts, and 13 and 14 as voltage ones (contact configuration R, R indicating `reversed', Fig. 9).
Note that the contact resistances were 0.5 and 0.8 $\Omega$ for contacts 15 and 16, respectively, which are smaller than 3.0 $\Omega$ for contact 14.
The first-harmonic resistance $R_1$ measured with the two contact configurations coincided with the maximum difference of only $\sim$0.2\% as expected for proper four-contact resistance measurements.
Unexpectedly, the antisymmetric part $R_2^a$ of the second-harmonic resistance measured with contact configuration R had the opposite sign to that measured with configuration N [compare Fig. 9(a) and Fig. 3(a)].
Also, the magnitude of $R_2$ was smaller.
The dc $R$ versus $I$ curve at $B$ = 0 is almost symmetric with respect to $I$ = 0, indicating a tiny linear-in-$I$ component [Fig. 9(b)].
The resistance increase with current is smaller than that in Fig. 5(a), indicating smaller joule heating.
A second-order polynomial fit (grey broken curve) gives $R_2^s$ = -0.0004 $\Omega$/A and that $R_3^s$ = 1.5 $\Omega$/A$^2$.
The former is much smaller than the value of -0.013 $\Omega$/A obtained with contact configuration N [Fig. 5(a)]. 
From the latter, the increase in the average sample temperature at $I$ = 3 mA is estimated to be 0.02 K, three times smaller than that with contact configuration N.
The $R$ versus $I$ curves at $B$ = $\pm$14 T indicate the existence of a linear-in-$I$ term, but the slope of $R^a$ is opposite to that obtained with contact configuration N [compare Fig. 9(c) and Fig. 5(b)], which is consistent with the ac measurements [Fig. 9(a)].
A linear fit to $R^a$ (grey broken line) gives $R_2^a$ = -0.041 $\Omega$/A, which is in good agreement with the ac measurements.
The symmetric (wrt $B$) part $R_s$ is almost symmetric with respect to $I$ = 0, and the resistance increase with current is small, similar to that at $B$ = 0.
From $R_s$ at $B$ = 14 T, the average temperature increase is estimated to be 0.02 K at $I$ = 3 mA.

The fact that the antisymmetric second-harmonic resistance $R_2^a$ changed sign as the current and voltage contacts were exchanged is difficult to explain if the second-harmonic resistance was intrinsic to sample bulk.
However, it is understandable if the second-harmonic voltage is due to joule heating at a current contact (or contacts): the use of different contacts produces different temperature gradient, resulting in different second-harmonic voltage generation.
The smaller magnitude of the second-harmonic resistance with current contacts 15 and 16 (contact configuration R) is also understandable because their contact resistances were smaller than the contact resistance of 14.

\begin{figure}
\includegraphics[width=8.6cm]{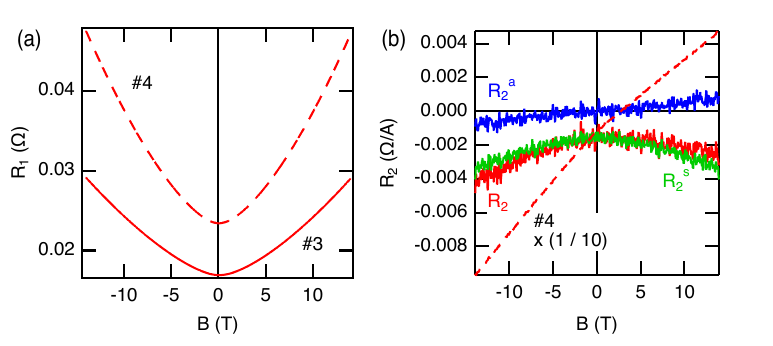}
\caption{\label{Sample}(a) First- and (b) second-harmonic resistance as a function of $B$ at $T$ = 30 K in sample \#3.  $R_2$ is decomposed into $R_2^s$ and $R_2^a$.  The magnetic field direction is the same as that in Fig. 3.  For comparison, $R_1$ and $R_2$ in sample \#4 with configuration N are shown with broken lines.  Note that $R_2$ for sample \#4 is scaled down by 1/10.
}
\end{figure}

\subsection{sample \#3}
We now present results for sample \#3, which show that the second-harmonic resistance can be negligibly small when contacts resistances are small.
Figure 10 shows the first- and second-harmonic resistances as a function of $B$ measured on sample \#3 at $T$ = 30 K.
The contacts 9 and 10 were used as current contacts (configuration N).
Both contact resistances in sample \#3 were small (0.3 and 0.7 $\Omega$) (Table I), in contrast to the fact that the contact resistance of contact 14 in sample \#4 was as large as 3.0 $\Omega$.
Figure 10 also shows $R_1$ and $R_2$ in sample \#4 with contact configuration N (broken lines), for comparison.
Note that $R_2$ in sample \#4 is scaled down by a factor of 1/10.
Although $R_1$ is of similar magnitude between samples \#3 and 4, $R_2$ is much smaller in sample \#3 than in \#4.
The parameter $\gamma^{\prime}$ at $B$ = 14 T is 3.5$\times 10^{-11}$ m$^2$A$^{-1}$T$^{-1}$ in sample \#3, more than one order-of-magnitude smaller than 8.8$\times 10^{-10}$ m$^2$A$^{-1}$T$^{-1}$ in sample \#4 with contact configuration N.

\begin{figure}
\includegraphics[width=8.6cm]{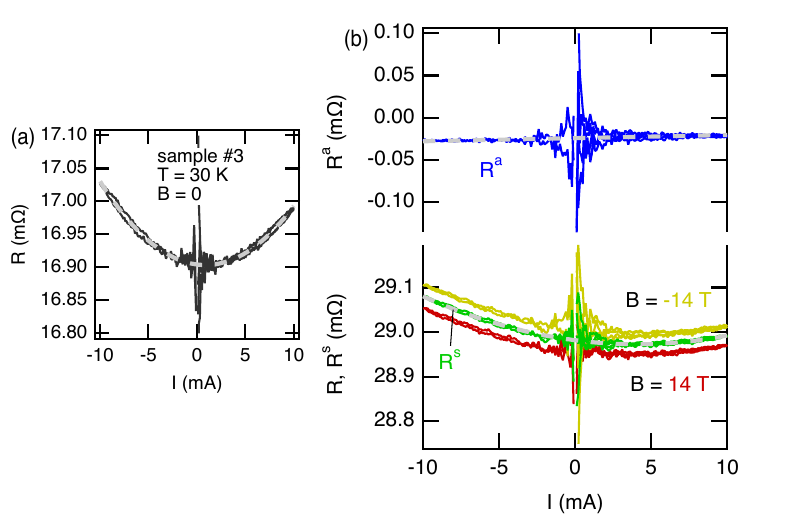}
\caption{\label{Sample}Dc $I-V$ measurements on sample \#3 at $T$ = 30 K and (a) at $B$ = 0 and (b) $\pm$14 T plotted in the form of $R$ versus $I$.
The magnetic field direction is the same as that in Fig. 3.
In (b), the symmetric (wrt. $B$) $R^s$ and the antisymmetric component $R^a$ are also shown.
Second-order polynomial fits to $R(I)$ at $B$ = 0 (a) and $R^s(I)$ at $B$ = 14 T (b) and a linear fit to $R^a(I)$ at $B$ = 14 T (b) are shown by grey broken lines.
}
\end{figure}

Figure 11 shows results of dc $I-V$ measurements on sample \#3.
Compared to the corresponding curve in Fig. 5(a), the $R$ versus $I$ curve at $B$ = 0 in Fig. 11(a) shows much less resistance change with increasing current and also is much less asymmetric with respect to $I$ = 0.
The second-order polynomial fit (grey broken line) gives $R_2^s$ = -0.0019 $\Omega$/A and $R_3^s$ = 1.1 $\Omega$/A$^2$.
The former is consistent with the ac measurements [Fig. 10(b)].
Using the latter value combined with a separately measured d$R$/d$T$ = 0.53 m$\Omega$/K, the increase in the average sample temperature at $I$ = 3 mA is estimated to be 0.02~K, much smaller than the value estimated for sample \#4 with contact configuration N (0.06 K).
Figure 11(b) shows the $R$ versus $I$ curves at $B$ = $\pm$14 T, together with the symmetric (wrt $B$) $R^s$ and the antisymmetric part $R^a$.
A second-order polynomial fit to $R^s$ at $B$ = 14 T [grey broken line in Fig. 11(b)] yields $R_2^s$ = -0.0044 $\Omega$/A and $R_3^s$ = 0.54 $\Omega$/A$^2$.
The former is in reasonable agreement with the ac measurements at $B$ = 14 T [Fig. 10(b)], and the latter with d$R$/d$T$ = 0.25 m$\Omega$/K at $B$ = 14 T gives the estimated increase in the average sample temperature at $I$ = 3 mA of 0.02 K, again much smaller than the corresponding value for sample \#4 (0.08 K).
The antisymmetric part $R^a$ is almost flat.
A linear fit to $R^a$ at $B$ = 14 T [grey broken line in Fig. 11(b)] gives a tiny value of $R_2^a$ = 0.00036 $\Omega$/A, which is in reasonable agreement with the ac measurements at $B$ = 14 T [Fig. 10(b)].
To summarize, the ac and dc measurements on sample \#3 indicated that joule heating of sample \#3 was much less than that of sample \#4 with contact configuration N and that the second-harmonic resistances $R_2^s$ and $R_2^a$ were correspondingly smaller. 

\begin{figure}
\includegraphics[width=8.6cm]{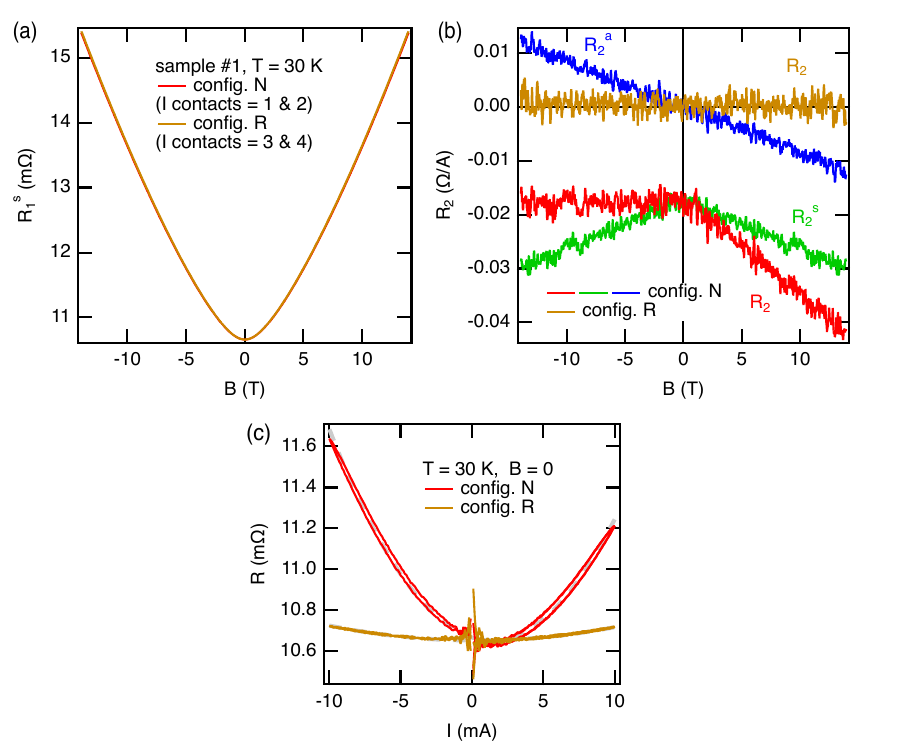}
\caption{\label{Sample}Sample \#1 with two different contact configurations N and R: The current contacts were 1 and 2 in configuration N, while 3 and 4 in R.  (a) First- and (b) second-harmonic resistance as a function of $B$ at $T$ = 30 K for the two configurations.  The magnetic field direction is the same as that in Fig. 3.  In (b), only $R_2$ in configuration N is decomposed into $R_2^s$ and $R_2^a$.  (c) dc $R$ versus $I$ curves at $T$ = 30 K and $B$ = 0 for the two configurations.
}
\end{figure}

\subsection{sample \#1}
Finally, we show results obtained for sample \#1, which will confirm that the nonzero $R_2^s$ and $R_2^a$ that we observed in bulk FeSe crystals were due to joule heating at current contacts and the thermoelectric effect.
Figure 12 shows results obtained with two contact configurations: 
In configuration N, contacts 1 and 2 were used as current contacts while 3 and 4 voltage ones.
In configuration R, contacts 3 and 4 were used as current contacts while 1 and 2 voltage ones.
Note that the contact resistance 2 was as large as 4.7 $\Omega$, whereas the rest of the contact resistances were small (Table I).

Figure 12(a) shows the first-harmonic resistances $R_1$ for the two configurations measured at $T$ = 30 K as a function of $B$.
The resistances $R_1$ for the two configurations were the same within experimental accuracy.
Figure 12(b) shows the second-harmonic resistances $R_2$ for the two configurations and the symmetric and antisymmetric parts for configuration N.
The second-harmonic resistances $R_2$ were clearly different between the two configurations:
$R_2$ for contact configuration N was finite, roughly of similar magnitude to that in sample \#4 with contact configuration N [Fig. 3(a)], whereas $R_2$ for configuration R is practically zero, below the noise level.
Figure 12(c) shows dc $R$ versus $I$ curves for the two configurations.
The resistance increase with current is much larger for configuration N than R, indicating larger joule heating for N.
A second-order polynomial fit (grey broken lines) gives
$R_2^s$ = -0.022 $\Omega$/A and $R_3^s$ = 8.0 $\Omega$/A$^2$ for configuration N, and
$R_2^s$ = -0.00038 $\Omega$/A and $R_3^s$ = 0.68 $\Omega$/A$^2$ for R.
$R_2^s$ is much larger for configuration N than for R, consistent with the ac results [Fig. 12(b)].
$R_3^s$ is also much larger for configuration N.
Using d$R$/d$T$ = 0.31 m$\Omega$/K from a separate measurement, the increase in the average sample temperature at $I$ = 3 mA is estimated to be 0.23 and 0.02 K for configuration N and R, respectively.
Thus, it is confirmed that the observation of the second-harmonic resistance is correlated to the joule heating at a current contact.

\section{Discussion}
We observed a rough but clear correlation between joule heating of samples and appearance of second-harmonic resistance:
A substantial second-harmonic resistance was observed in sample \#4 with configuration N and sample \#1 with configuration N (Figs. 3 and 12), for which the estimated average sample temperature increase at $I$ = 3 mA was 0.06 and 0.23 K, respectively.
In contrast, the second-harmonic resistance was vanishingly small in sample \#3 with configuration N and sample \#1 with configuration R (Figs. 10 and 12), for which the temperature increase was smaller, 0.02 K.
It is however to be noted that, because the temperature gradient and resulting thermoelectric voltage depends not only on the magnitude of joule heating but also on details of heat exchange between the sample and environment, no simple relation between the average temperature increase and the magnitude of the second-harmonic resistance is expected.
This explains the observation that a significant second-harmonic resistance was observed in sample \#4 with configuration R despite the estimated temperature increase of 0.02 K (Fig. 9).
It is also to be noted that the local temperature increase at a current contact is expected to be much larger than the above mentioned average temperature increases.
It is this local temperature increase that governs the magnitude of the thermoelectric voltage.

The thermoelectric coefficients in FeSe can be found in the literature such as refs. \cite{McQueen79PRB, Kasahara16NatCommun, Yang17PRB, Chen20PRB}.
The Seebeck coefficient $S$ is positive at room temperature and changes sign near 230 K.
Although it shows a broad negative peak near 100 K, the temperature dependence below 150 K is weak, $S$ being of the order of $-$10~$\mu$VK$^{-1}$, and the anomaly associated with the structural transition at $T_s$ is subtle \cite{McQueen79PRB}.
The Nernst coefficient $\nu$ reported in \cite{Yang17PRB} gradually increases from $\sim$150 K peaks near 80 K, $\nu$ being $\sim$0.9~$\mu$VK$^{-1}$T$^{-1}$.
The anomaly at $T_s$ is unclear.
These reports are consistent with the present observations that the second-harmonic resistances $R_2^s$ and $R_2^a$ started to appear far above $T_s$ and that almost no  anomaly was observed at $T_s$ (Fig. 2).
The magnitudes of the thermoelectric coefficients show substantial sample dependence.
One to two orders-of-magnitude larger values of $S$ and $\nu$ than those in \cite{McQueen79PRB, Yang17PRB} were reported in \cite{Kasahara16NatCommun}.
This is likely related with the fact that the sample of \cite{Kasahara16NatCommun} shows much larger magnetoresistance than that of \cite{Yang17PRB}.
The magnetoresistance of the present samples is also much larger than that of \cite{Yang17PRB}, and hence the magnitudes of their thermoelectric coefficients are likely close to those in \cite{Kasahara16NatCommun}.

Let us make an order-of-magnitude estimate of thermoelectric coefficients necessary to explain the observed $R_2^s$ and $R_2^a$.
Let us assume a typical size of the second-harmonic resistance to be 0.02 $\Omega$/A.
This corresponds to a voltage of $\sim$0.2~$\mu$V at $I$ = 3 mA.
Let us assume the temperature difference between the voltage contacts of 1 K.
Then, a Seebeck coefficient of 0.2~$\mu$VK$^{-1}$ suffices.
For the Nernst effect, the estimation is more difficult because the direction of the temperature gradient and the geometry of contacts are involved.
For simplicity, we consider a temperature difference of 1 K in some direction in the $ab$ plane.
A Nernst coefficient of 2~$\mu$VK$^{-1}$T$^{-1}$ gives a voltage of 20~$\mu$V in the perpendicular direction at $B$ = 10 T applied along the $c$ axis.
Only 1\% of this voltage is necessary to produce $R_2^a$ = 0.02 $\Omega$/A.
Considering large thermoelectric coefficients reported in \cite{Kasahara16NatCommun}, the above values of $S$ and $\nu$ are reasonable assumptions.
Thus, our claim that the observed second-harmonic resistance is actually due to the thermoelectric effect is justified.

The unexpectedly large influence of the thermoelectric effect on the second-harmonic resistance is due to the large thermoelectric coefficients in FeSe \cite{Kasahara16NatCommun}.
Our results clearly support ref. \cite{Nagata24condmat}, in which the authors observed the zero-field superconducting diode effect in FeSe devices and ascribed it to the thermoelectric effect.
The implications of the present results are twofold:
Firstly, the thermoelectric effect can be used to design superconducting diodes.
By using superconductors with large thermoelectric effect like FeSe and intentionally making temperature gradient in a device, effective superconducting diodes may be achieved as in \cite{Nagata24condmat}.
Secondly, caution has  to be taken when using the second-harmonic resistance to make a diagnosis of broken space-inversion symmetry.
The appearance of the antisymmetric second-order harmonic resistance is sometimes taken as evidence for broken space-inversion symmetry.
However, such diagnosis has to be made with extreme caution to exclude possible contamination by the thermoelectric effect.
This is especially true when dealing with semimetals or semiconductors with small Fermi energy because those materials tend to have large thermoelectric coefficients \cite{Behnia16RPP}.

\begin{acknowledgments}
This work was supported by Grant-in-Aid for Scientific Research on Innovative Areas ``Quantum Liquid Crystals'' (No. JP19H05824, JP22H04485), Grant-in-Aid for Scientific Research(A) (No. JP21H04443, JP22H00105, JP23H00089), Grant-in-Aid for Scientific Research(B) (No. JP22H01173), Grant-in-Aid for Scientific Research(C) (No. JP22K03537), and Fund for the Promotion of Joint International Research (No. JP22KK0036) from Japan Society for the Promotion of Science.
MANA is supported by World Premier International Research Center Initiative (WPI), MEXT, Japan.
\end{acknowledgments}

\end{document}